\documentclass[showpacs,10pt,twocolumn,prb]{revtex4}
\usepackage{amsmath}
\usepackage{amssymb}
\usepackage{graphics}
\usepackage{epsfig}
\usepackage{color}

\begin{document}

\title{Phase diagram of superconductivity and antiferromagnetism in single
crystals of Sr(Fe$_{1-x}$Co$_{x}$)$_{2}$As$_{2}$ and Sr$_{1-y}$Eu$_{y}$(Fe$%
_{0.88}$Co$_{0.12}$)$_{2}$As$_{2}$}
\author{Rongwei Hu, Sergey L. Bud'ko, Warren E. Straszheim, Paul C. Canfield}
\affiliation{Ames Laboratory, U.S. DOE and Department of Physics and Astronomy, Iowa
State University, Ames, IA 50011, USA}
\date{\today }

\begin{abstract}
We report magnetic susceptibility, resistivity and heat capacity
measurements on single crystals of the Sr(Fe$_{1-x}$Co$_{x}$)$_{2}$As$_{2}$
and Sr$_{1-y}$Eu$_{y}$(Fe$_{0.88}$Co$_{0.12}$)$_{2}$As$_{2}$ series. The
optimal Co concentration for superconductivity in Sr(Fe$_{1-x}$Co$_{x}$)$%
_{2} $As$_{2}$ is determined to be $x\sim 0.12$. Based on this we grew
members of the Sr$_{1-y}$Eu$_{y}$(Fe$_{0.88}$Co$_{0.12}$)$_{2}$As$_{2}$
series so as to examine the effects of well defined, local magnetic moments,
on the superconducting state. We show that superconductivity is gradually
suppressed by paramagnetic Eu$^{2+}$ doping and coexists with
antiferromagnetic ordering of Eu$^{2+}$ as long as $T_{c}>T_{N}$. For $y\geq
0.65$, $T_{N}$ crosses $T_{c}$ and the superconducting ground state (as
manifested by zero resistivity) abruptly disappears with evidence for
competition between superconductivity and local moment antiferromagnetism
for $y$ up to 0.72. It is speculated that the suppression of the
antiferromagnetic fluctuations of Fe sublattice by coupling to the long
range order of Eu$^{2+}$ sublattice destroys bulk superconductivity when $%
T_{N}>T_{c}$.
\end{abstract}

\pacs{74.25.Dw, 74.25.Fy, 74.25.Ha, 74.62.Dh}
\maketitle

\section{Introduction}

The interplay between superconductivity (SC) and magnetism has been of a
long standing interest in condensed matter physics. SC and magnetism were
originally considered to be mutually exclusive in conventional
superconductors because magnetism breaks the time reversal symmetry of the
singlet Cooper pairs. The influence of paramagnetic impurities on SC was
first studied theoretically by Abrikosov and Gor'kov (AG).\cite{Abrikosov}
It was shown that SC is drastically suppressed by dilute magnetic moments
due to the spin-flip scattering. Early experimental investigations were
limited to superconducting systems without long-range magnetic order.\cite%
{Matthias}$^{-}$\cite{Williams} The coexistence of SC and long-range
magnetism was realized in several families of ternary and quaternary
rare-earth compounds discovered later, also referred to as magnetic
superconductors, e.g. RMo$_{6}$S$_{8}$ , RRh$_{4}$B$_{4}$ and RNi$_{2}$B$%
_{2} $C.\cite{Machida}$^{-}$\cite{Canfield2} In these compounds, the
localized $4f $ electrons of the rare-earth ions are indirectly coupled via
conduction electrons by the Ruderman-Kittel-Kasuya-Yosida interaction (RKKY)
and responsible for various magnetic orderings. The conduction electrons,
often primarily from the transition metal, give rise to SC. The coexistence
is more favorable for antiferromagnetism (AF), since the AF molecular field
exerted on SC electrons may be averaged out on the scale of SC coherence
length.

Another type of magnetic superconductor is the one where the moment is
itinerant. In itinerant electron systems long range order may be carried by
the same electrons that become superconducting, leading to competition
(sometime strong) between the two states. The recently discovered iron
arsenic based superconductors appear to be one such example. The parent
compounds (RFeAsO 1111 series, R=La, Ce, Pr, Nd, Sm or Gd, and AFe$_{2}$As$%
_{2}$ 122 series, alkali earth A=Ca, Sr, Ba) are semimetals and show either
closely spaced, or a simultaneous AF ordering and tetragonal to orthorhombic
(ortho) structural transition. With electron or hole doping, the magnetic
and structural transitions are suppressed to low temperature and SC, with $%
T_{c}$ up to 55 K\cite{Kamihara}, is induced. Ba$_{1-x}$K$_{x}$Fe$_{2}$As$%
_{2}$ \cite{Rotter} exhibits a maximum $T_{c}$ of $37$ K, or by substitution
of transition metal for Fe, e.g. Ba(Fe$_{1-x}$Co$_{x}$)$_{2}$As$_{2}$, $%
T_{c} $ can reach 22 K\cite{Sefat}$^{-}$\cite{Ni1} or for Sr(Fe$_{1-x}$Co$%
_{x}$)$_{2}$As$_{2}$, $T_{c}$ can reach 18 K.\cite{Jasper} Unlike the 1111
series for which the magnetic/structural transition was suggested to
disappear abruptly prior to the emergence of SC, the 122 series show a
gradual suppression of the AF/ortho transition, which coexists with SC for a
range of dopings.\cite{Canfield3}$^{,}$\cite{Ni1} For the Co doped Ba-122
series neutron scattering shows a suppression of the magnetic order
parameter on entering the superconducting state, indicating strong coupling
between AF and SC.\cite{Canfield3}$^{,}$\cite{Pratt} In addition, both $\mu $%
SR\cite{Bernhard}$^{,}$\cite{Khas} and $^{75}$As NMR\cite{Laplace}%
measurements unambiguously indicate that AF order is present in all of the
sample volume when the sample is in the superconducting state, i.e. that the
magnetic order and SC coexist homogeneously at the atomic scale. Whereas
Fe-based AF coexists with SC and Fe based AF fluctuations may well be vital
to FeAs based superconductors, a systematic study of effects of well
defined, local magnetic moments on this SC is lacking. Starting from
optimally Co-doped SrFe$_{2}$As$_{2}$, we can have Eu$^{2+}$ substituting
for Sr$^{2+}$ without introducing extra electrons/holes and assess the
sensitivity of this SC to the large $J=S=7/2$ local moment.

The Eu end compound, EuFe$_{2}$As$_{2}$, exists as an isostructural member
of the 122 series. Therefore a continuous substitution can be expected
between EuFe$_{2}$As$_{2}$ and AFe$_{2}$As$_{2}$. EuFe$_{2}$As$_{2,}$ in
addition to the AF order of the iron sublattice at about 189 K, exhibits an
A-type AF order of Eu$^{2+}$ ions at 19 K.\cite{Jeevan} On suppression of
the AF order of iron with pressure or Co doping\cite{Miclea}$^{-}$\cite%
{Nicklas}, the onset of SC was observed, which was then followed by a
resistive reentrance attributed to the magnetic order of Eu$^{2+}$. Given
the sensitivity of the 122 compounds to strain/pressure\cite{Milton}$^{-}$%
\cite{Park}, we choose SrFe$_{2}$As$_{2}$ as host, owing to the similar size
of Sr$^{2+}$(118 pm) and Eu$^{2+}$ (117 pm)\cite{Shannon}, so as to minimize
the steric effects of\ the doping.

In order to perturb the SC of the Sr 122 phase by isoelectronic substitution
of Eu and establish phase diagrams systematically using the same growth
technique for Co and Eu doping, the phase diagram of Sr(Fe$_{1-x}$Co$_{x}$)$%
_{2}$As$_{2}$ as a function of Co substitution is constructed first. The
optimal Co doping level of $x\sim 0.12$ is then kept the same across the
whole range of Eu doping. We present the magnetic susceptibility,
resistivity and heat capacity measurements on Sr$_{1-y}$Eu$_{y}$(Fe$_{1-x}$Co%
$_{x}$)$_{2}$As$_{2}$. Superconductivity of the optimally Co doped SrFe$_{2}$%
As$_{2}$ is suppressed gradually by Eu doping ($0\leq y<0.43$), crosses over
a region with coexistence of SC and Eu based AF ($0.43\leq y\leq 0.60$) with 
$T_{N}$ increasing linearly with $y$. For $y\geq 0.65$, $T_{N}$ cuts across
the $T_{c}$ line and SC suddenly disappears leaving just the Eu$^{2+}$, AF
ordered state. An initial study of EuFe$_{2}$As$_{2}$ doped with both Sr and
Co was recently published, but using samples with nominal doping values and
focusing on the Eu-rich side.\cite{He} We will compare the results of our
systematic study with Ref. 32 in the discussion section.

\section{Experiment}

Single crystal samples of both Sr(Fe$_{1-x}$Co$_{x}$)$_{2}$As$_{2}$ and Sr$%
_{1-y}$Eu$_{y}$(Fe$_{0.88}$Co$_{0.12}$)$_{2}$As$_{2}$ were grown via a self
flux method.\cite{Ni1}$^{,}$\cite{Paul} The FeAs and CoAs precursors were
first synthesized by solid state reaction. Elemental Sr and Eu were mixed
with FeAs and CoAs in the stoichiometry of $1:4-4x:4x$ and $1-y:y:3.44:0.56$
respectively in an alumina crucible and sealed into an amorphous silica
tube. The sealed ampoule was heated to 1180 $^{\circ }$C and then cooled
slowly to 1000 $^{\circ }$C; finally the excess liquid flux was decanted.%
\cite{Paul} The as-grown crystals were annealed under a static Ar atmosphere
at 500 $^{\circ }$C for 24 hours (as discussed below in Section III).\cite%
{Saha}

\begin{figure}[tbp]
\centerline{\includegraphics[scale=0.35]{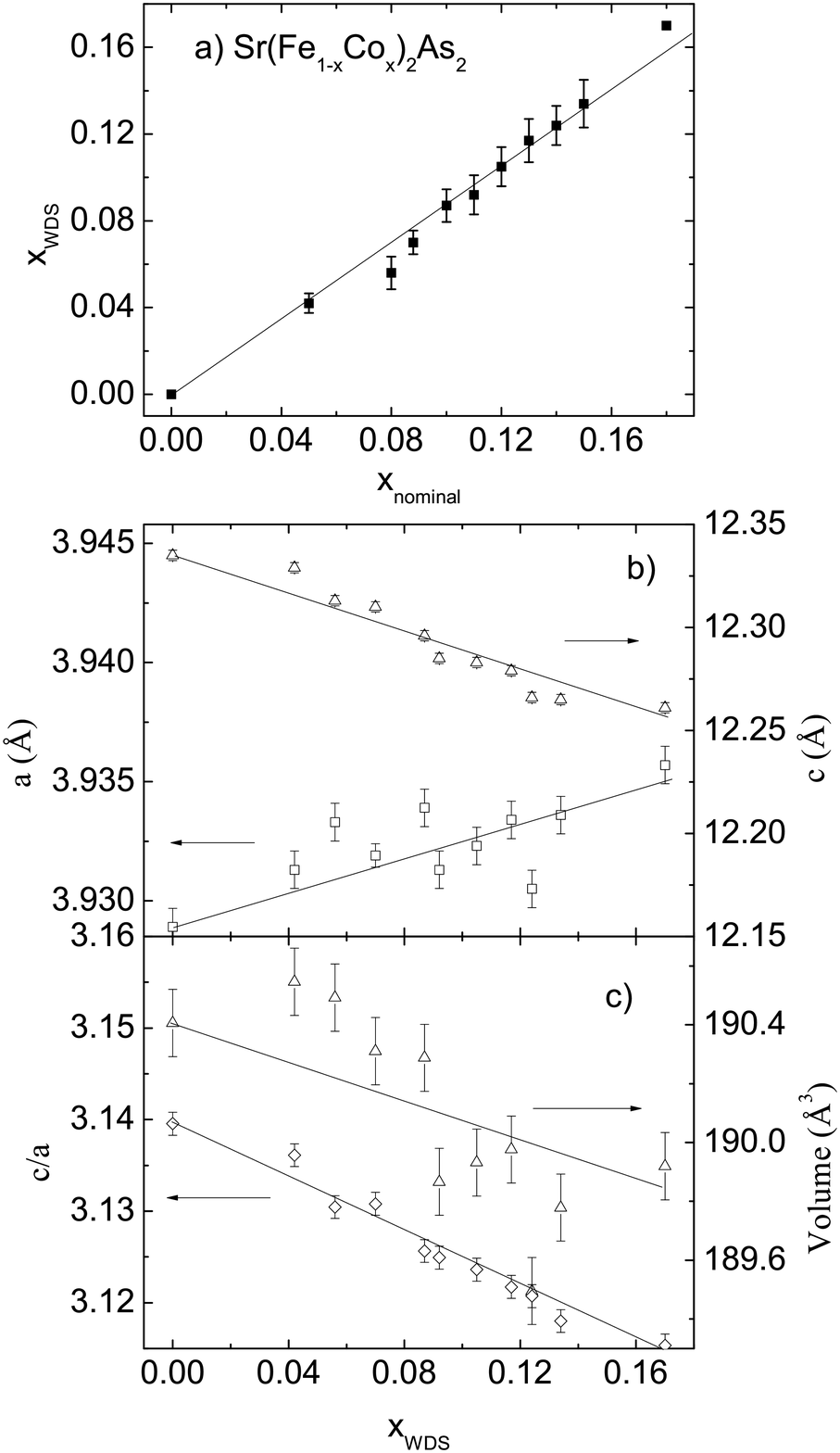}} \vspace*{-0.3cm}
\caption{Results of elemental analysis and lattice parameters determined on
Sr(Fe$_{1-x}$Co$_{x}$)$_{2}$As$_{2}$. a) Measured Co concentration from WDS
vs. nominal one; b-c) Lattice parameters, \textit{a}, \textit{c}, \textit{c/a%
} and unit cell volume as a function of $x_{WDS.}$}
\end{figure}

Powder x-ray diffraction, with Si standard, was performed using a Rigaku
Miniflex X-ray diffractometer with Cu K$\alpha $ radiation ($\lambda =1.5418%
\mathring{A}$). The lattice parameters were refined by Rietica software.\cite%
{Rietica} Chemical composition was determined by wavelength dispersive x-ray
spectroscopy (WDS) in a JEOL JXA-8200 electron microscope. The actual
composition of the single crystals was taken as the average of 10 spots
measured on the crystal and the error bar was taken as the standard
deviation of the 10 values.

Magnetic susceptibility was measured in a Quantum Design MPMS, SQUID
magnetometer. The in-plane AC resistivity was measured by a standard
four-probe method using an LR-700 resistance bridge\ with an excitation of
60 $\mu V$ on samples of typical size $3$ $mm\times 2$ $mm\times 0.2$ $mm$.
Electrical contacts were made using Dupont 4929N silver paint. Heat capacity
data were collected using a Quantum Design PPMS.

All the samples were found to slowly degrade in air. Over a period of four
months, a ferromagnetic background on the order of 10$^{-2}$ emu/mol
develops, although no obvious change in appearance of the crystal and no
impurity phase in powder XRD pattern can be observed. Elemental analysis
indicates significant\ presence of oxygen in the surface layer of the aged
samples, implying the formation of oxides. In addition, the superconducting
transition of the aged sample broadens and T$_{c}$ decreases as measured by
low field magnetization. Therefore all the measurements reported in this
paper were performed shortly after the samples were prepared. It should be
noted that whereas the Sr-based 122 compounds are known to be susceptible to
chemical changes\cite{Hidenori}, as well as strain\cite{Saha1}, we observed
no sample quality change over time in the well studied\cite{Canfield3}$^{,}$%
\cite{Ni1} Ba(Fe$_{1-x}$Co$_{x}$)$_{2}$As$_{2}$ samples.

\begin{figure}[tbp]
\centerline{\includegraphics[scale=0.35]{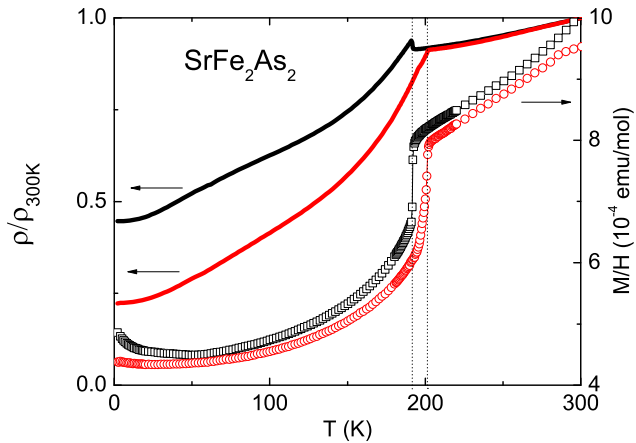}} \vspace*{-0.3cm}
\caption{The annealing effect for pure SrFe$_{2}$As$_{2}$. As-grown sample
(black), annealed sample (red)}
\end{figure}

\begin{figure}[tbp]
\centerline{\includegraphics[scale=0.6]{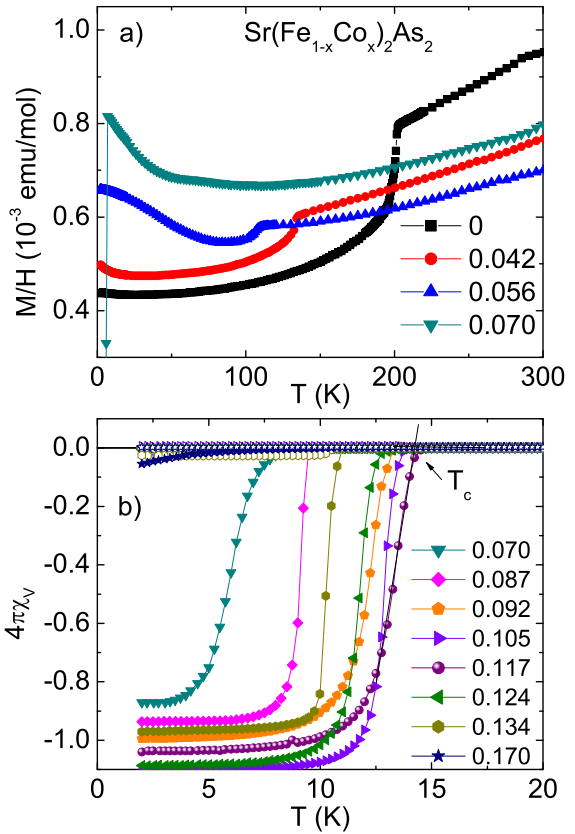}} \vspace*{-0.3cm}
\caption{a) Magnetic susceptibility as a function of temperature for Sr(Fe$%
_{1-x}$Co$_{x}$)$_{2}$As$_{2}$ single crystals taken at 10 kOe with H$\Vert $%
\textit{ab}; b) Low field (100 Oe) magnetic susceptibility. Field-cooled
curves are shown in open symbols. T$_{c}$ is infered from the intersect of
the steepest slope to the normal magnetic suscpetibility.}
\end{figure}

\begin{figure}[tbp]
\centerline{\includegraphics[scale=0.35]{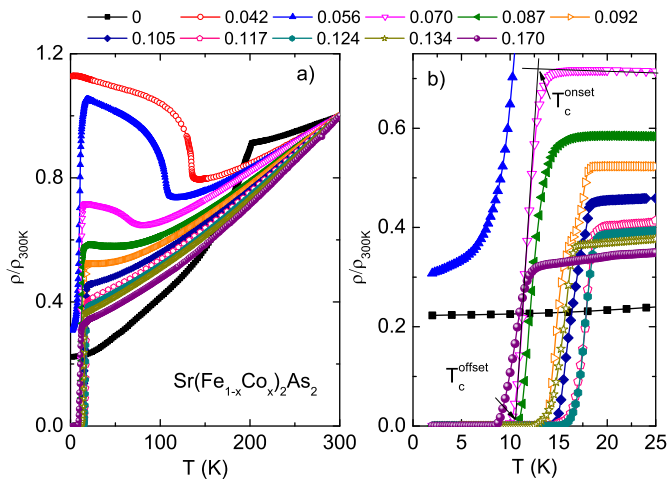}} \vspace*{-0.3cm}
\caption{a) Resistivity normalized to the room temperature value $\protect%
\rho (T)/\protect\rho (300K)$ for Sr(Fe$_{1-x}$Co$_{x}$)$_{2}$As$_{2}$ ($%
0\leq x\leq 0.17$); b) Low temperature resistivity showing superconducting
transtion.}
\end{figure}

\section{Results and discussions}

\subsubsection{Sr(Fe$_{1-x}$Co$_{x}$)$_{2}$As$_{2}$}

The results of elemental analysis and lattice parameter determined from the
powder X-ray measurements on Sr(Fe$_{1-x}$Co$_{x}$)$_{2}$As$_{2}$ are shown
in Fig. 1. The nearly linear dependence in Fig. 1(a), with a slope of 0.94,
indicates good agreement between the actual Co concentration, $x_{WDS}$, and
the nominal concentration, $x_{nominal}$. The compositional spread over a
wide area on the sample surface for each concentration is less than 0.02.
These results demonstrate the relative homogeneity of the Co doping in the
single crystal samples. Figures 1(b) and (c) show that the lattice
parameters \textit{a} and\textit{\ c}, as well as the \textit{c/a }ratio and
unit cell volume as a function of $x_{WDS}$. The parameter \textit{c}, and 
\textit{c/a},\textit{\ }change linearly with $x_{WDS}$ and the values are in
good agreement with the previous report.\cite{Jasper} By substitution of Co
for Fe the lattice is changed more along \textit{c} axis than in the \textit{%
ab} plane. The lattice parameter \textit{c} decreases by 0.6\% ($0.074%
\mathring{A}$) for $x=0.17$, whereas the lattice parameter \textit{a}
increases by only about 0.2\% ($0.007\mathring{A}$). The random error of
lattice parameter determined by our Miniflex X-ray diffractometer is about
0.02\%, $\sim 0.0008\mathring{A}$ for the \textit{a }lattice parameter,
which is about the same order as the average deviation from a linear
variation $\sim 0.0013\mathring{A},$\textit{. }Thus lattice parameters can
be regarded to vary linearly with $x_{WDS}$, within experimental errors, in
accordance with Vegard's law.

Annealing can have clear effect on samples and has been shown to remove
extrinsic effects associated with strain induced defects.\cite{Saha} As
shown in Fig. 2, the magnetic/structural transition of pure SrFe$_{2}$As$%
_{2} $ is increased from 192 K, for the as grown sample to 201 K, for the
annealed sample, which is very close to the previous reported values for
polycrystalline (205 K)\cite{Jasche} and Sn flux-grown single crystalline
(198 K)\cite{JQ Yan} SrFe$_{2}$As$_{2}$. Based on these observations, our
samples are heat treated under the conditions described in Section II above.

The magnetic susceptibility for H $\Vert $ \textit{ab} of the Sr(Fe$_{1-x}$Co%
$_{x}$)$_{2}$As$_{2}$ series was measured in a magnetic field of 10 kOe for $%
x\leq 0.07$ (Fig. 3 (a)). The parent compound SrFe$_{2}$As$_{2}$ manifests a
sharp drop at 201 K in magnetic susceptibility, due to the
magnetic/structural transition.\cite{Jasche}$^{,}$\cite{JQ Yan} With
increasing Co doping, this transition is suppressed to lower temperature and
becomes undetectable for $x>0.07$. For $0.07\leq x\leq 0.17$, SC is induced
and is manifested in low field (H = 100 Oe) zero-field-cooled (ZFC) and
field-cooled (FC) measurements below 20 K (Fig. 3 (b)). The data are
compared to 1/4$\pi $ to give a rough estimate of the superconducting volume
fraction. Although, as discussed in Ref. 39, the FC curves are routinely
close to zero in these materials, due to pinning or surface barrier effects,
the ZFC curves approaching -1 suggest bulk SC. The superconducting
transitions remain very sharp for $x\geq 0.07,$ except for $x=0.17$ which
becomes broad and barely visible, consistent with a $T_{c}$ reduced to a
value close to our base temperature. The transition temperature increases
from 7.4 K for $x=0.07$, maximizes at 14.8 K for $x=0.117$ and then
diminishes to 5.7 K for $x=0.17$.

Fig. 4(a) shows the temperature dependence of the electrical resistivity of
Sr(Fe$_{1-x}$Co$_{x}$)$_{2}$As$_{2}$, normalized to the room temperature
values. Similar to the case of Ba(Fe$_{1-x}$Co$_{x}$)$_{2}$As$_{2}$ series%
\cite{Ni1}, the magnetic/structural transition of Sr(Fe$_{1-x}$Co$_{x}$)$%
_{2} $As$_{2}$\ manifests itself as a sudden drop for $x=0$ and as an
increase in resistivity for $x=0.042-0.07$ and nearly disappears for $%
x=0.087 $. After the magnetic/structural transition is completely suppressed
for $x\geq 0.092$, the series shows featureless, metallic temperature
dependence. Fig. 4(b) shows an expanded view for low temperatures. At $%
x=0.056$, a broad and incomplete superconducting transition is observed;
zero resistance is only reached for $x\geq 0.07$, this agrees with the bulk
SC observed in magnetic measurements.

In order to establish the phase diagram for the Sr(Fe$_{1-x}$Co$_{x}$)$_{2}$%
As$_{2}$ series, the transition temperatures were inferred in the same
manner as used in Ref. 16. $T_{c}$ from magnetic susceptibility is
determined from the intersection of the steepest slope and the linear
extrapolation of normal magnetic susceptibility, shown in Fig. 3 (b).
Resistive onset and offset of $T_{c}$ values are inferred from the
intersects of the steepest slope with the normal state and zero resistance
respectively, shown in Fig. 4 (b). $T_{M/S}$ is inferred from the peak of $%
d(M/H)/dT$ and $d[\rho /\rho (300K)]/dT$; data for $x=0.056$ is shown in
Fig. 5 as an example. It is argued by Gillett \textit{et al}\cite{Gillett}
that only a single, first-order-like, transition occurs in the heat capacity
of Sr(Fe$_{1-x}$Co$_{x}$)$_{2}$As$_{2}$ with coincidence of magnetic and
structural transitions. Our magnetization and resistance data also do not
show a discernible splitting between $T_{M}$ and $T_{S}$, Fig. 5, further
supports this observation.

Based on our magnetization and electrical resistance measurements, the phase
diagram of Sr(Fe$_{1-x}$Co$_{x}$)$_{2}$As$_{2}$ is mapped out in Fig. 6. A
superconducting dome is found: SC is first stabilized for $x=0.07$ at about
7.4 K, reaches a maximum $T_{c}$ of $\sim $14.5 K for $x=0.117$, then
decreases to 5.7 K for $x=0.17$. Our phase diagram is in good agreement with
earlier ones. The phase diagram for polycrystalline Sr(Fe$_{1-x}$Co$_{x}$)$%
_{2}$As$_{2}$ showed a complete suppression of magnetic/structural
transition and appearance of SC at $x_{nominal}=0.1$ with the highest $T_{c}$
of 19 K.\cite{Jasper} The difference between maximum $T_{c}$ of the
polycrystalline and our single crystalline samples is probably due to strain
effect. As it has been demonstrated\cite{Milton}$^{-}$\cite{Park}, strain
can affect Sr122 profoundly, especially when there is a high surface area to
volume fraction (as in powders). Results consistent with our single crystal
Sr(Fe$_{1-x}$Co$_{x}$)$_{2}$As$_{2}$ work, with highest $T_{c}\sim 13$ K
were reported by Kasinathan \textit{et al}.\cite{Deepa} The more recent one
based on self-flux grown single crystals, having larger density of data
points, showed more clearly a coexistence of $T_{M/S\text{ }}$ and SC
transition for $x=0.07\sim 0.09$ and the superconducting dome with optimal $%
T_{c}$ of 16 K at $x=0.10$.\cite{Gillett} The differences between our phase
diagram and the published ones, in terms of transition temperature and
optimal doping concentration, may be associated with differences in both
sample preparation and uncertainties of concentration. For our self-flux
grown samples, we can choose Co concentration $x\sim 0.12$, with the highest 
$T_{c}$ and suppressed AF/ortho transition as the starting point for our
study of the effects of local moments of FeAs based superconductor via Eu
substitution for Sr.

\subsubsection{Sr$_{1-y}$Eu$_{y}$(Fe$_{1-x}$Co$_{x}$)$_{2}$As$_{2}$}

For our Sr$_{1-y}$Eu$_{y}$(Fe$_{1-x}$Co$_{x}$)$_{2}$As$_{2}$ series, the Co
concentration was kept at $x\sim 0.12$ and the series was doped by Eu for $%
0\leq y\leq 1$. Fig. 7(a) shows the elemental analysis results for the
actual Eu and Co concentrations as a function of nominal Eu concentration.
The actual concentration of Eu agrees well with the nominal, with a slope of
1.03, and the Co concentration is essentially constant. The lattice
parameters \textit{a}, \textit{c} and unit cell volume are plotted in Figs.
7(b) and (c). Compared to Sr(Fe$_{0.883}$Co$_{0.117}$)$_{2}$As$_{2}$ ($%
a=3.9334(2)$ $\mathring{A},$ $c=12.2790(2)$ $\mathring{A}$), the smaller Eu$%
^{2+}$ ion leads to a decrease in \textit{c} axis by 2\% ($0.256\mathring{A}$%
) and a decrease in \textit{a} axis by 0.4\% ($0.014\mathring{A}$). The
small concentration error and linear dependence on $x_{WDS}$ indicate a
homogeneous substitution of Sr by Eu across the whole series. 
\begin{figure}[tbp]
\centerline{\includegraphics[scale=0.33]{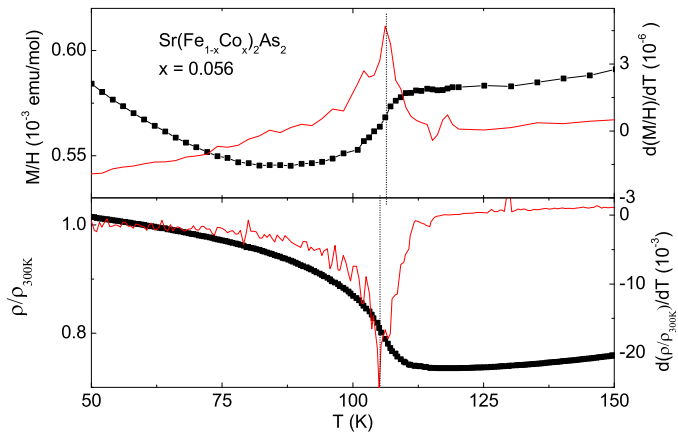}} \vspace*{-0.3cm}
\caption{Magnetic susceptibility, normalized resistivity and the temperature
derivatives, single peak associated with simutaneous magnetic and structural
transition.}
\end{figure}
\begin{figure}[tbp]
\centerline{\includegraphics[scale=0.35]{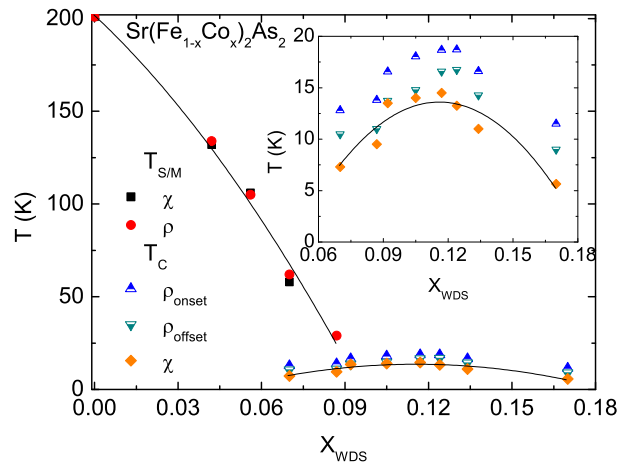}} \vspace*{-0.3cm}
\caption{Temperature and chemical composition phase diagram of Sr(Fe$_{1-x}$%
Co$_{x}$)$_{2}$As$_{2}$ single crystals for $0\leq x\leq 0.17$. Lines are
guide to the eye.}
\end{figure}
\begin{figure}[tbp]
\centerline{\includegraphics[scale=0.35]{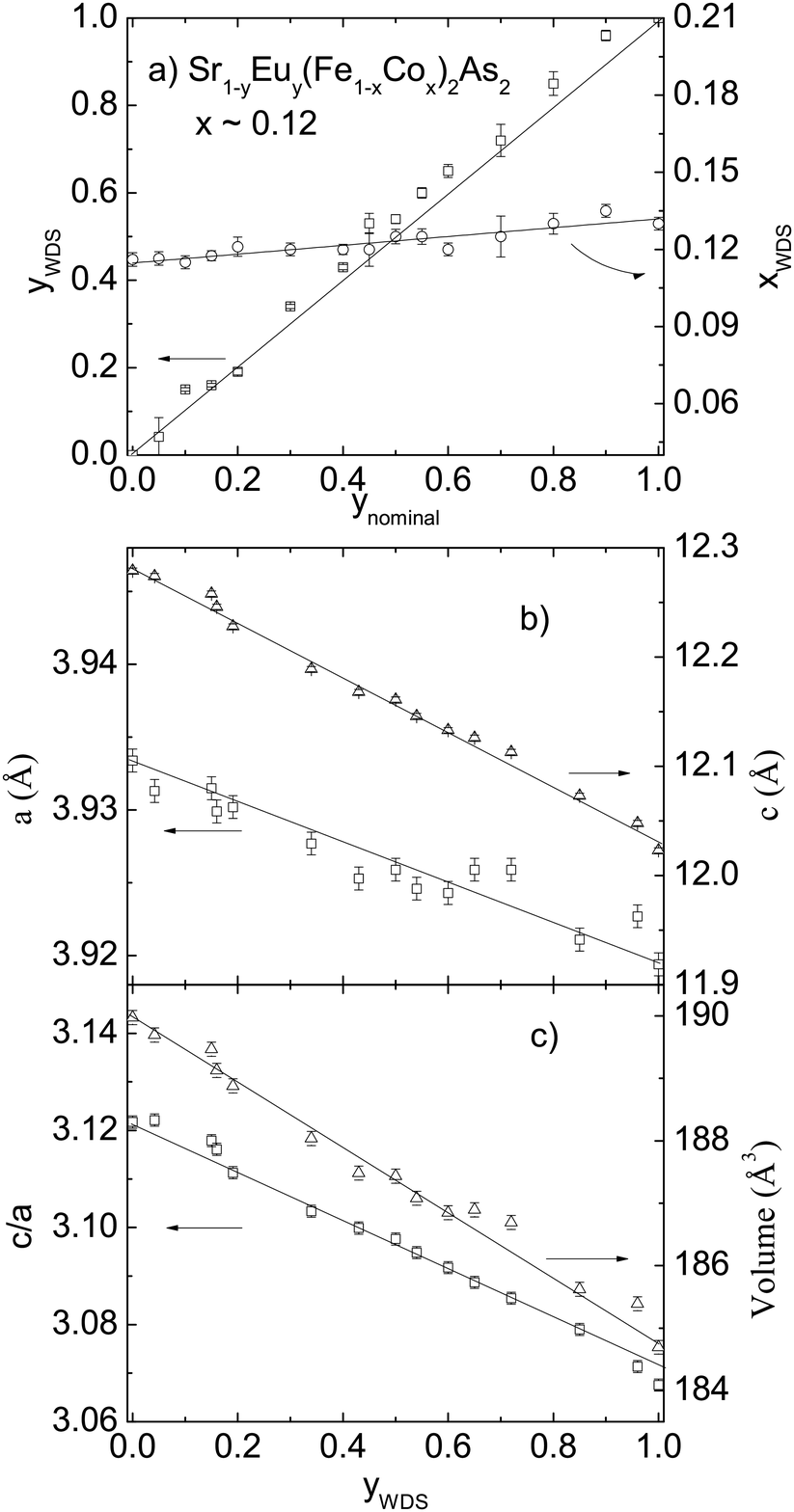}} \vspace*{-0.3cm}
\caption{a) Elemental analysis of Sr$_{1-y}$Eu$_{y}$(Fe$_{0.88}$Co$_{0.12}$)$%
_{2}$As$_{2};$ b)-c) Lattice parameters, a, c and c/a, as well as unit cell
volume.}
\end{figure}

\begin{figure}[tbp]
\centerline{\includegraphics[scale=0.65]{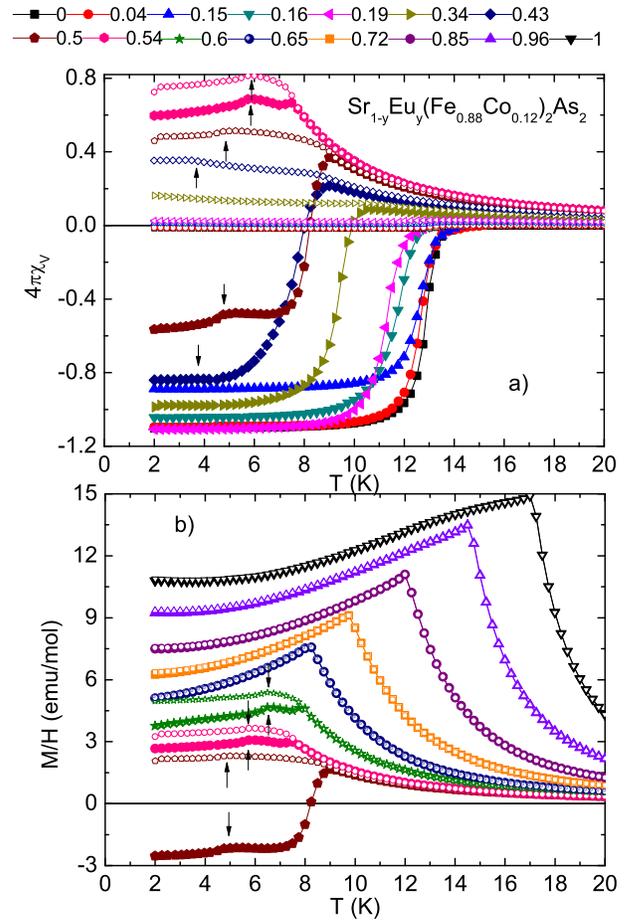}} \vspace*{-0.3cm}
\caption{Magnetic susceptibility of Sr$_{1-y}$Eu$_{y}$(Fe$_{0.88}$Co$_{0.12}$%
)As$_{2}$ single crystals taken for 100 Oe magnetic field applied within the 
\textit{ab} plane. Solid symbols denote ZFC data and open symbols denote FC
data. Arrows indicate the AF transitions, which are consistent for both ZFC
and FC curves.}
\end{figure}

\begin{figure}[tbp]
\centerline{\includegraphics[scale=0.75]{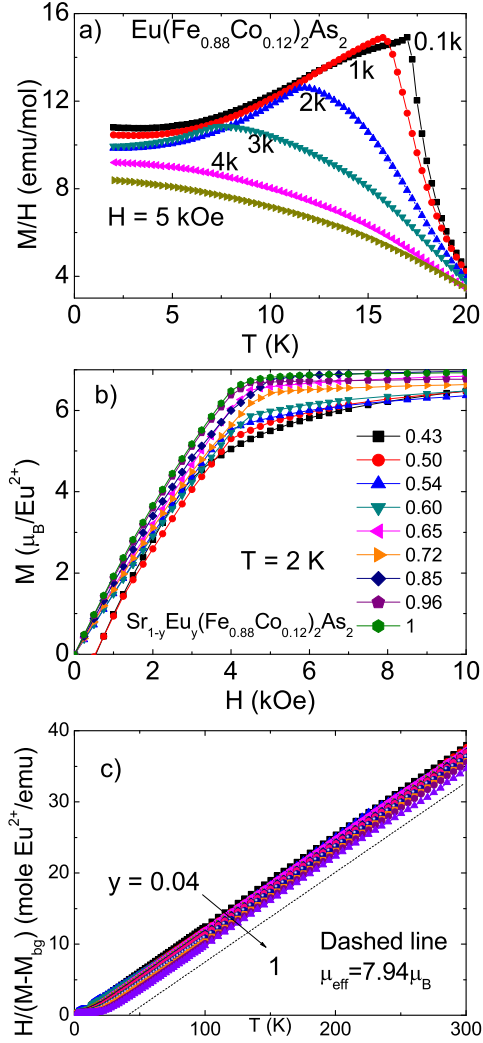}} \vspace*{-0.3cm}
\caption{a) Field dependence of the temperature dependent magnetization
divided by applied field for $y=1$ for H $\Vert $ \textit{ab} plane; b)M(H)
normalized by $y_{WDS}$; c) Inverse magnetic susceptibility, H/M normalized
by the actual Eu$^{2+}$ concentration $y_{WD\dot{S}\text{.}}$}
\end{figure}

\begin{table*}[tph]
\caption{Results of elemental analysis for Sr$_{1-y}$Eu$_{y}$(Fe$_{1-x}$Co$%
_{x}$)$_{2}$As$_{2}$ and the Eu$^{2+}$ concentration inferred from high
temperature magnetic susceptibility. The Curie-Weiss temperature $\protect%
\theta _{CW}$ is compared with the AFM transition temperature $T_{N}$.}
\label{tbl1}%
\begin{tabular}{l|llllllllllllll}
\hline\hline
$y_{nominal}$ & $0.05$ & $0.1$ & $0.15$ & $0.2$ & $0.3$ & $0.4$ & $0.45$ & $%
0.5$ & $0.55$ & $0.6$ & $0.7$ & $0.8$ & $0.9$ & $1$ \\ \hline
$y_{WDS}$ & $0.04$ & $0.15$ & $0.16$ & $0.19$ & $0.34$ & $0.43$ & $0.50$ & $%
0.54$ & $0.60$ & $0.65$ & $0.72$ & $0.85$ & $0.96$ & $1$ \\ 
$y_{M}$ & $0.05$ & $0.12$ & $0.15$ & $0.18$ & $0.31$ & $0.40$ & $0.48$ & $%
0.50$ & $0.60$ & $0.64$ & $0.73$ & $0.85$ & $0.97$ & $1$ \\ 
$\theta _{CW}(K)$ & $2.2$ & $4.1$ & $4.0$ & $4.2$ & $5.0$ & $9.3$ & $9.7$ & $%
10.3$ & $11.7$ & $13.9$ & $14.6$ & $18.6$ & $19.4$ & $20.2$ \\ 
$T_{N}(K)$ &  &  &  &  &  & $3.5$ & $4.5$ & $5.5$ & $6.3$ & $8.0$ & $9.5$ & $%
11.8$ & $14.3$ & $16.8$ \\ \hline\hline
\end{tabular}%
\end{table*}

The in-plane magnetic susceptibility of Sr$_{1-y}$Eu$_{y}$(Fe$_{0.88}$Co$%
_{12}$)$_{2}$As$_{2}$ is shown in Fig. 8. Both ZFC and FC curves are
measured in a magnetic field of 100 Oe. The data clearly indicate that there
are three regions of low temperature behavior across the series: i) $0\leq
y\leq 0.34$, SC is gradually suppressed by Eu doping but remains a simply
identifiable transition; ii) $0.43\leq y\leq 0.60,$ in this intermediate
range, the Curie-Weiss paramagnetic background due to Eu$^{2+}$ moments
gradually becomes large enough to shift the diamagnetic signal to positive
values. In addition a second feature appears and as $y$ increases it rises
in temperature leading to a double-peak feature, which can be ascribed to
the coexistence of SC and lower temperature AF associated with the Eu$^{2+}$
sublattice. The upper transition shows a splitting between ZFC and FC curves
consistent with SC. The lower transition of AF origin is present on both ZFC
and FC curves at the same temperature, indicated by arrows. These
transitions are further confirmed by heat capacity measurement as shown
below; iii) $0.65\leq y\leq 1$, clear AF transitions manifest as cusps and $%
T_{N}$ continues to increase with Eu$^{2+}$ doping up to 17 K for $y=1$. It
is worth noting that FC and ZFC curves collapse on each other for these
higher $y$ values, suggesting long range antiferromagnetic order, similar to
EuFe$_{2}$As$_{2}$\cite{Ren}, instead of other magnetic origin, e.g. spin
glass or ferrimagnetic order.

Fig. 9(a) shows M(T)/H as a function of temperature data measured in various
fields for $y=1$. The cusp initially shifts to lower temperature with higher
field and then becomes saturated paramagnetic-like for fields above 4 kOe.\
Neutron scattering experiments on pure EuFe$_{2}$As$_{2}$ revealed that the
long range order of Eu$^{2+}$ is of A-type AF, namely the Eu$^{2+}$ moments
are parallel in \textit{ab} plane and antiparallel along \textit{c} axis
with an ordering wavevector of $k=(0,0,1)$.\cite{Xiao} Therefore the
meta-magnetic transition for $y=1$ is most likely due to the spin flip along
the field direction between Eu$^{2+}$ layers, similar to EuFe$_{2}$As$_{2}$.%
\cite{Jiang} Our results are in good agreement with the reported magnetic
field dependence of M/H for EuFe$_{1.715}$Co$_{0.285}$As$_{2}$, where
meta-magnetic transition occurs at a lower field of 3.5 kOe than that of
pure EuFe$_{2}$As$_{2}$ (8.5 kOe).\cite{Jiang} Because of this meta-magnetic
transition, the series for $y\geq 0.43$ all show similar field dependence
(Fig. 9(b)), i.e. the slope of magnetization changes around 4 kOe and shows
a saturation moment of $\sim $7$\mu _{B}$/Eu$^{2+}$ in high field. For $%
y=0.43$ and $0.50$, diamagnetic contribution of SC can be seen below 500 Oe.
Given the meta-magnetic transition, the AF transition temperature $T_{N\text{
}}$was inferred from the cusp of $d(\chi T)/dT$ measured in $H=100$ $Oe$.%
\cite{Fisher1}

In Fig. 9(c) we examine the high temperature behavior of the magnetic
susceptibility. Since the Hund's rule ground state for Eu$^{2+}$ is the same
as Gd$^{3+}$ ($^{7/2}S$), there is no spin-orbital coupling and thus the
crystal field effect is absent and well defined magnetic moments of Eu$^{2+}$
exhibiting Curie-Weiss law at high temperatures are expected. We are able to
estimate the concentration of Eu$^{2+}$ from magnetic measurements by
assuming each Eu$^{2+}$ carries an effective magnetic moment of 7.94$\mu
_{B} $. The magnetic background of Sr(Fe$_{0.883}$Co$_{0.117}$)$_{2}$As$_{2}$
in a magnetic field of 10 kOe is subtracted from all the datasets and the
inverse magnetic susceptibility normalized to a fitted Eu concentration $%
y_{M}$ is plotted in Fig. 9(c) as a function of temperature. The magnetic
susceptibility above 100 K is fitted by the Curie-Weiss law:

\begin{equation*}
\chi (T)=\frac{y_{M}N\mu (Eu^{2+})^{2}}{3k_{B}(T-\theta _{CW})}=\frac{%
7.94^{2}y_{M}}{8(T-\theta _{CW})}[emu/mol]
\end{equation*}%
where N is Avogadro constant, $k_{B}$ is the Boltzmann constant and $\theta
_{CW}$ is Curie-Weiss temperature. As can be seen in Table I, $y_{WDS}$ and $%
y_{M}$ agree well with each other and follow the same trend with the nominal
concentration. The positive Curie-Weiss temperature is consistent with an
overall predisposition to ferromagnetic coupling between Eu$^{2+}$ moments,
at least in the magnetic field of 10kOe.

The low temperature ( $T<20K$) heat capacity divided by temperature, $%
C_{p}/T $, vs $T$ of Sr$_{1-y}$Eu$_{y}$(Fe$_{0.88}$Co$_{0.12}$)$_{2}$As$_{2}$
is presented in Fig. 10(a). A very pronounced discontinuity can be seen for $%
0.43\leq y\leq 1$. The transition temperature $T_{N}$, defined by this
discontinuity, decreases with decreasing Eu$^{2+}$ concentration and is in
excellent agreement with the cusp of $d(\chi T)/dT$ of magnetic
susceptibility. These data confirm that AF is the lower transition in the
intermediate range $0.43\leq y\leq 0.60$. For the $y=0.34$ data this
discontinuity appears to be at or just below our base temperature of 2.0 K.
For $y<0.34$ the complete transition can not be detected. It is worth noting
that the low temperature $C_{p}/T$ below 5 K for $0.65\leq y\leq 1$ show a
linear dependence on T, i.e. $C\propto T^{2}$. This temperature dependence
of heat capacity is consistent with the low temperature AF magnon
excitations of a two dimensional magnetic lattice.\cite{Jongh}

\begin{figure}[tbp]
\centerline{\includegraphics[scale=0.65]{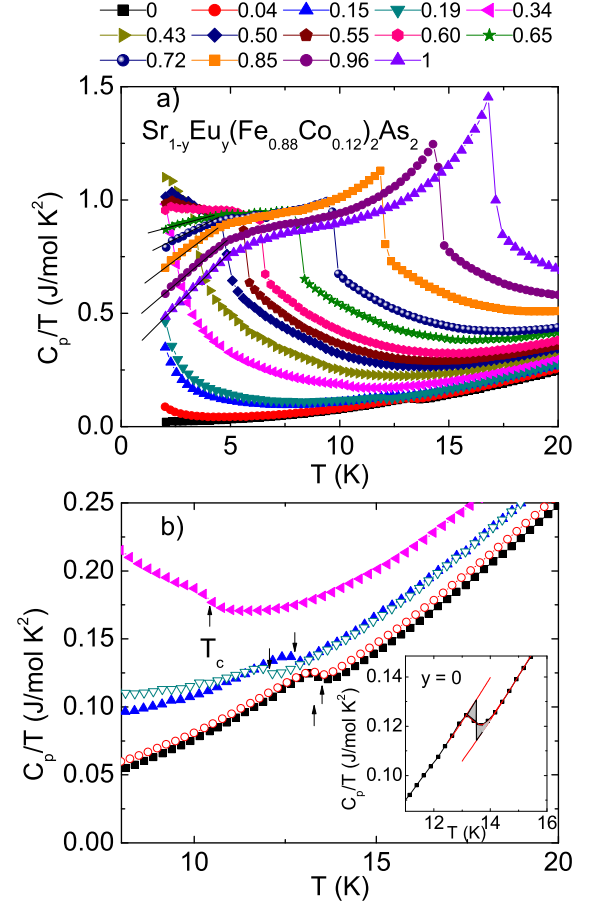}} \vspace*{-0.3cm}
\caption{a) $C_{p}/T$ vs. $T$ of Sr$_{1-y}$Eu$_{y}$(Fe$_{1-x}$Co$_{x}$)$_{2}$%
As$_{2}$, solid line indicates $C_{p}\sim T^{2}$ for a FM magnon
contribution; b)an expanded view for $0\leq y\leq 0.19$ showing a SC jump, $%
T_{c}$ is indicated by arrows. Inset shows the isoentropic reconstruction of
the superconducting transition of $C_{p}/T.$}
\end{figure}

Figure 10(b) shows that starting from the low $y$ side, SC can be identified
as a weak jump for $0\leq y\leq 0.19$, but becomes hard to detect for $%
y>0.34 $ because of the large background associated with the AF transition.
Fig. 10(b) inset shows a representative heat capacity jump for $y=0$. The SC
transition temperature $T_{c}$ is inferred by isoentropic construction, i.e.
the two shaded areas have the same size. For $y=0.34$, $T_{c}$ is taken as
the middle point of the jump.

\begin{figure}[tbp]
\centerline{\includegraphics[scale=0.8]{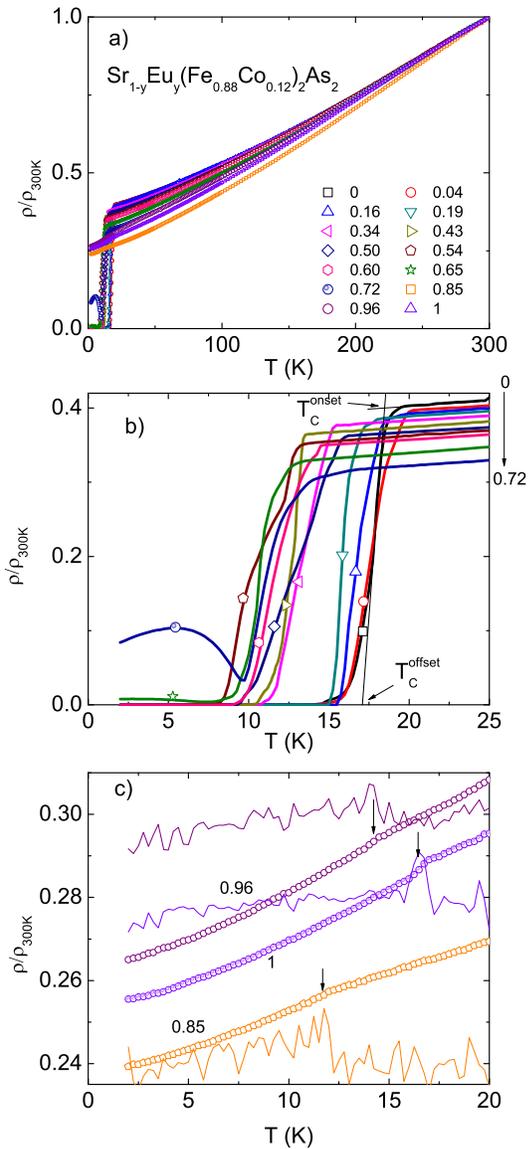}} \vspace*{-0.3cm}
\caption{a) Normalized electrical resistivity of Sr$_{1-y}$Eu$_{y}$(Fe$%
_{0.88}$Co$_{0.12}$)$_{2}$As$_{2}$; b) Low temperature data showing the
superconducting transition; c) For $0.85\leq y\leq 1$, the loss of spin
scattering around $T_{N}$, solid lines are the temperature derivatives.}
\end{figure}

\begin{figure}[tbp]
\centerline{\includegraphics[scale=0.4]{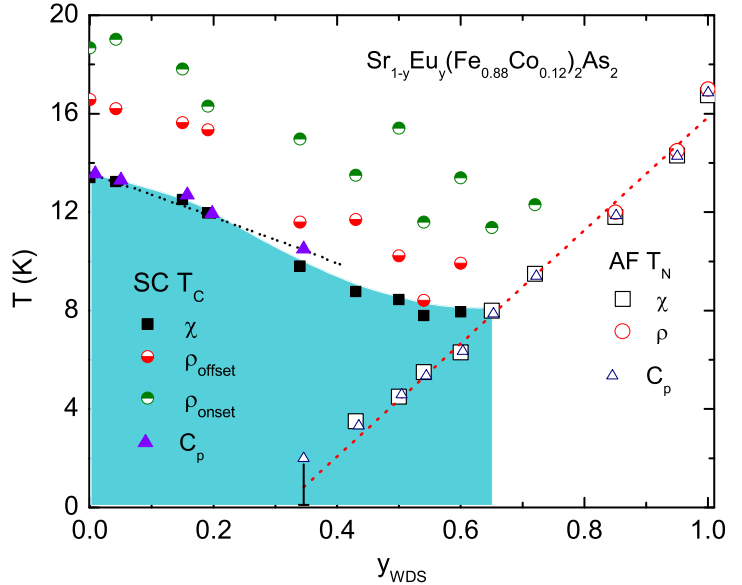}} \vspace*{-0.3cm}
\caption{$T-y$ phase diagram of Sr$_{1-y}$Eu$_{y}$(Fe$_{0.88}$Co$_{0.12}$)$%
_{2}$As$_{2}$ single crystals. Black dashed line is the fit to AG theory.
Red dashed line highlights $T_{N}(y)$ and is just a guide to the eye. Blue area is the superconducting region.}
\end{figure}

Figure 11 shows the temperature dependence of the normalized resistivity of
Sr$_{1-y}$Eu$_{y}$(Fe$_{0.88}$Co$_{12}$)$_{2}$As$_{2}$. Given the Co-doping
level (near optimal), it is not surprising that the series remains metallic
and featureless above 20 K. In Fig. 11(b) it can be seen that
superconducting transition temperature is gradually lowered by Eu$^{2+}$
doping for $0\leq y\leq 0.60$. The transition becomes broader for $y>0.19,$
e.g. $\Delta T$ is 4 K for $y=0$ and 6 K for $y=0.5$. This wide transition
is similarly observed in Ni doped SrFe$_{2}$As$_{2}$.\cite{Saha} For $y=0.65$
and $0.72$, a resistivity reentrance is observed as a broad peak below a
local minimum in resistivity at 7.8 and 9.8 K respectively. The minimum
coincides with the AF order temperature measured by magnetic susceptibility
and heat capacity, indicating that the bulk SC transition is interrupted by
AF order. Such incomplete resistive transitions have been observed in Sr$%
_{0.3}$Eu$_{0.7}$(Fe$_{0.86}$Co$_{0.14}$)$_{2}$As$_{2}$ and EuFe$_{2}$As$%
_{2} $ under pressure,\cite{Miclea}$^{,}$\cite{He}

The superconducting transition temperature, $T_{c}$, is inferred in the same
way as in Fig. 4. For $y=0.65$ and $0.72$, only $T_{c}^{onset}$ is
extracted. For $0.85\leq y\leq 1$, Fig. 11(c), the series remains a normal
metal and manifests a very small change in slope at $T_{N}$ (corresponding
to the peak in $d[\rho /\rho (300K)]/dT$), due to the loss of spin disorder
scattering. We must note that the change in resistivity at ${T}_{N}$ is very
small, even smaller than that of EuFe$_{2}$As$_{2}$.\cite{Ren} It implies
very weak coupling between Eu$^{2+}$ moments and conduction electrons. A
recent detailed transport studies of EuFe$_{2}$As$_{2}$ under high pressure
showed that electron scattering due to Eu$^{2+}$ has minor contribution to
both resistivity and Hall effect, thus consistent with our conclusion.\cite%
{Taichi1}

\subsubsection{Analysis and discussion}

Based on the transport and thermodynamic measurements, a phase diagram as a
function of Eu doping can be constructed and is shown in Fig. 12. Starting
from the Eu-rich side of the phase diagram, we can see that $T_{N}$
decreases with decreasing Eu content and crosses through the $T_{c}$ line,
near $y\sim 0.60$, without any resolvable change in slope ($dT_{N}/dy$).
This is fairly standard behavior for an intermetallic compound with a local
moment antiferromagnetic phase transition that is being reduced via site
dilution with a non-magnetic ion (i.e. Sr$^{2+}$ for Eu$^{2+}$).\cite%
{Canfield2}$^{,}$\cite{Wiener} Starting from the Sr-rich side of the phase
diagram we can see that when Eu$^{2+}$ is a paramagnetic impurity, it
suppresses SC monotonically, but rather weakly. The weakness of the
paramagnetic Eu$^{2+}$ as a pair breaker is not unexpected, given the rather
weak coupling of the Eu$^{2+}$ moments to the conduction electrons, as most
clearly manifested by the small loss of spin-disorder scattering seen in
Fig. 11(c). The suppression of $T_{c}$ by magnetic impurities in a
nonmagnetic superconductor has been discussed by Abrikosov and Gor'kov.\cite%
{Abrikosov} The fit to AG theory for data $0\leq y\leq 0.34$ gives a
critical concentration $y_{c}=1.08$, implying SC could survive in Eu(Fe$%
_{0.88}$Co$_{0.12}$)$_{2}$As$_{2}$ if the Eu$^{2+}$ sublattice were to
remain in the disordered paramagnetic state (\textit{which it does not}).
For $0.43\leq y\leq 0.60$, both SC and AF states are clearly detected. As
long as $T_{c}>T_{N}$, the advent of AF order does not lead to any
re-entrance or other clear features in the $T-y_{WDS}$ phase diagram. This
is in agreement with early findings that EuFe$_{2}$As$_{2\text{ }}$becomes a
bulk superconductor with $T_{c}\sim 30$ K and SC coexists with AFM order
with $T_{N}\sim 20$ K.\cite{Taichi2} The remarkable feature revealed in Fig.
12 is the sudden disappearance of bulk SC when the $T_{N}$ line intercepts
the $T_{c}$ line. Superconductivity, as defined by a $\rho =0$ state,
suddenly disappears for $y\geq 0.65$. This sudden truncation of the
superconducting region is quite remarkable and demands further analysis.

As has been shown in this work and discussed before\cite{Saha1}$^{,}$\cite%
{Saha}, the resistivity data associated with pure and doped SrFe$_{2}$As$%
_{2} $ samples is complicated, manifesting superconducting transition
temperatures that appear to be higher than those determined by bulk,
thermodynamic measurements such as magnetic susceptibility and specific
heat. On the other hand, in both figures 6 and 12 the superconducting
transition inferred from resistivity roughly tracks those inferred from
magnetization and specific heat (in Fig. 12 even the $T_{c}$ data inferred
from onset criterion drop by a similar amount as the $T_{c}$ values inferred
from thermodynamic data, just with an offset by a few degrees). This is
consistent with the idea that a small portion of the sample has an enhance $%
T_{c}$ associated with some strain/damage. Similar difference of $T_{c}$
between that inferred from resistivity and magnetic susceptibility, as well
as transition width $\Delta T_{c}$, were observed in Fig. 6 and in Sr(Fe$%
_{1-x}$Ni$_{x}$)$_{2}$As$_{2}$\cite{Saha} This being said, the absence of
any hint of superconducting drop in the $y>0.72$ data (Fig. 11 (c)) is a
conclusive evidence that there is not even trace superconductivity in these
samples. For the $y=0.65$ and $0.72$ samples there appears to be an onset of
filamentary superconductivity that is interrupted by the bulk AF.

In comparison to Ref. 32, whereas the above discussion and data show that AF
appears to be very detrimental to the formation of the superconducting state
when $T_{N}>T_{c}$, there is no evidence of the AF leading to dramatic
re-entrance of the normal state when $T_{N}<T_{c}$ (i.e. for $y\leq 0.60$).
The resistivity data, as well as the susceptibility data do not show any
feature that can be associated with the re-establishment of the normal state
below the $T_{N}$ line as it cuts under the superconducting state.

These observations have several implications and also suggest several
directions for future research. First, although dilute, paramagnetic, Eu$%
^{2+}$ only weakly suppresses SC, antiferromagnetically ordered Eu$^{2+}$
appears to prevent its formation. As has been the case for other magnetic
superconductors, specifically the RNi$_{2}$B$_{2}$C materials\cite{Canfield2}%
, a dramatic difference in the effects of local moments on SC can be
observed when comparing disordered, single ions in paramagnetic state, and
an antiferromagnetically ordered sublattice. In the case of (Ho$_{1-x}$Dy$%
_{x}$)Ni$_{2}$B$_{2}$C\cite{Canfield2}$^{,}$\cite{Cho} as $T_{c}$ crosses
from above $T_{N}$ to below it, the cause of pair breaking changes from
spin-flip scattering off of single impurities to interactions with magnetic
excitations of the order state. In the case of Sr$_{1-y}$Eu$_{y}$(Fe$_{0.88}$%
Co$_{0.12}$)$_{2}$As$_{2}$ the sudden loss of superconductivity as $T_{N}$
rises above $T_{c}$ implies that somehow long range antiferromagnetic order
of the Eu sublattice strongly suppresses (or removes) necessary ingredients
for the establishment of the superconducting state. If antiferromagnetic
fluctuations of the Fe-sublattice (associated with the $k=(1,0,1)$ ordering%
\cite{Zhao}) are associated with the pairing in the superconducting state,
then long range order of the large ($J=S=7/2$) Eu sublattice with an
ordering wave vector of $k=(1,0,0)$ could easily be related to a dramatic
change in the Fe sublattice fluctuation spectrum. Such a dramatic change in
the fluctuations could easily be the suppressed, missing ingredient for
superconductivity invoked above. So, unlike DyNi$_{2}$B$_{2}$C, which
apparently requires antiferromagnetic ordering of the Dy sublattice to
suppress pair breaking of the individual Dy moments\cite{Canfield2}, Sr$%
_{1-y}$Eu$_{y}$(Fe$_{0.88}$Co$_{0.12}$)$_{2}$As$_{2}$ requires the Eu
sublattice to remain in the disordered, paramagnetic state in order to
establish the FeAs-based superconducting state.

Although this hypothesis readily explains the sudden loss of SC when $%
T_{N}>T_{c}$, it also would imply that the SC state below $T_{N}$, when $%
T_{N}<T_{c}$, should be modified; although figures 8, 11, and 12 show that
there is no effect of $T_{N}$ on the low field magnetization and zero field
resistivity when $T_{N}<T_{c}$, it is reasonable to anticipate that there
will be changes in other superconducting parameters such as the superfluid
density and penetration depth.

\section{Conclusions}

Transport and thermodynamic measurements were performed on Sr(Fe$_{1-x}$Co$%
_{x}$)$_{2}$As$_{2}$ and Sr$_{1-y}$Eu$_{y}$(Fe$_{0.88}$Co$_{0.12}$)$_{2}$As$%
_{2}$ single crystals. A superconducting dome is identified in Sr(Fe$_{1-x}$%
Co$_{x}$)$_{2}$As$_{2}$ as a function of Co doping and the optimal Co
concentration is determined to be $x\sim 0.12$. The SC of the optimal Co
doping is gradually suppressed by paramagnetic Eu$^{2+}$ following AG theory
and found to coexist with AF of Eu$^{2+}$ for $0.43\leq y\leq 0.60$. For
higher Eu$^{2+}$ doping, bulk SC disappears suddenly when $T_{N}>T_{c}$. We
speculate that the long range order of Eu$^{2+}$ sublattice is coupled to
the AF fluctuations of Fe sublattice and the suppression of the Fe
fluctuations required for FeAs-based SC is what gives rise to the abrupt
loss of bulk SC when $T_{N}$ surpasses $T_{c}$.

\section{Acknowledgements}

The authors acknowledge Alex Thaler for experimental assistance, Cedomir
Petrovic, J\"{o}rg Schmalian and Rafael M. Fernandes for helpful
discussions. This work was carried out at the Iowa State University and
supported by the AFOSR-MURI grant \#FA9550-09-1-0603 (Rongwei Hu, Paul C.
Canfield). Part of this work was performed at Ames Laboratory, US DOE, under
contract \# DE-AC02-07CH 11358 (Sergey L. Bud'ko, Warren E. Straszheim and
Paul C. Canfield). Sergey L. Bud'ko was also partially supported by the
State of Iowa through the Iowa State University.


\bigskip

\begin{thebibliography}{99}
\bibitem{Abrikosov} A.A. Abrikosov, L.P. Gor'kov, Sov. Phys. JETP \textbf{16}%
, 1575 (1962)

\bibitem{Matthias} B.T. Matthias, H. Suhl, E. Corenzwit, Phys. Rev. Lett, 
\textbf{1}, 92 (1958)

\bibitem{Mathhias2} Matthias B T, Suhl H and Corenzwit E, Phys. Rev. Lett. 
\textbf{1} 449 (1958)

\bibitem{Williams} L. J. Williams, W. R. Decker, and D. K. Finnemore, Phys.
Rev. B \textbf{2}, 1287 (1970)

\bibitem{Machida} K. Machida, Appl. Phys. A \textbf{35, }193 (1984)

\bibitem{Muller} K-H M\"{u}ller and V N Narozhnyi, Rep. Prog. Phys. \textbf{%
64} 943 (2001).

\bibitem{Gupta} L. C. Gupta, Advances in Physics, \textbf{55}, 691 (2006).

\bibitem{Fisher} O \O . Fischer, M. B. Maple, Superconductivity in Ternary
Compounds I, Structural, Electronic and Lattice Properties, Springer (1982).

\bibitem{Buschow} K. Buschow, E. Wohlfarth, Ferromagnetic Materials, Chap.
6, Elsevier, Amsterdam (1990).

\bibitem{Canfield} Charles P. Poole Jr., Paul. C. Canfield, Arthur P.
Ramirez, Handbook of superconductivity, Pages 71-108 (2000).

\bibitem{Canfield2} Paul C. Canfield, Peter L. Gammel and David J. Bishop,
Physics Today, \textbf{51}, 40 (1998).

\bibitem{Kamihara} Y. Kamihara, T. Watanabe, M. Hirano, and H. Hosono, J.
Am. Chem. Soc. \textbf{130}, 3296 (2008).

\bibitem{Rotter} M. Rotter, M. Tegel, and D. Johrendt, Phys. Rev. Lett. 
\textbf{101}, 107006 (2008).

\bibitem{Sefat} A. S. Sefat, R. Jin, M. A. McGuire, B. C. Sales, D. J.
Singh, and D. Mandrus, Phys. Rev. Lett. \textbf{101}, 117004 (2008).

\bibitem{Canfield3} Paul C. Canfield and Sergey L. Bud'ko, Annual Review of
Condensed Matter Physics, 1, 27 (2010).

\bibitem{Ni1} N. Ni, M. E. Tillman, J.-Q. Yan, A. Kracher, S. T. Hannahs, S.
L. Bud'{}ko, and P. C. Canfield, Phys. Rev. B \textbf{78,} 214515 (2008).

\bibitem{Jasper} A. Leithe-Jasper, W. Schnelle, C. Geibel, and H. Rosner,
Phys. Rev. Lett. \textbf{101}, 207004 (2008).

\bibitem{Pratt} D. K. Pratt, W. Tian, A. Kreyssig, J. L. Zarestky, S. Nandi,
N. Ni, S. L. Bud'ko, P. C. Canfield, A. I. Goldman, and R. J. McQueeney,
Phys. Rev. Lett. \textbf{103}, 087001 (2009).

\bibitem{Bernhard} C Bernhard, A J Drew, L Schulz, V K Malik, M Rossle, Ch
Niedermayer, Th Wolf, G D Varma, G Mu, H-H Wen, H Liu, G Wu and X H Chen,
New J. Phys., 11, 055050 (2009).

\bibitem{Khas} R. Khasanov, A. Maisuradze, H. Maeter, A. Kwadrin, H.
Luetkens, A. Amato, W. Schnelle, H. Rosner, A. Leithe-Jasper, and H.-H.
Klauss, Phys. Rev. Lett. 103, 067010 (2009).

\bibitem{Laplace} Y. Laplace, J. Bobroff, F. Rullier-Albenque, D. Colson,
and A. Forget, Phys. Rev. B 80, 140501 (2009).

\bibitem{Jeevan} H. S. Jeevan, Z. Hossain, Deepa Kasinathan, H. Rosner, C.
Geibel and P. Gegenwart, Phys. Rev. B \textbf{78}, 052502 (2008).

\bibitem{Miclea} C. F. Miclea, M. Nicklas, H. S. Jeevan, D. Kasinathan, Z.
Hossain, H. Rosner, P. Gegenwart, C. Geibel and F. Steglich, Phys. Rev. B 
\textbf{79}, 212509 (2009).

\bibitem{Nicklas} M. Nicklas, M. Kumar, E. Lengyel, W. Schnelle, A.
Leithe-Jasper, arXiv:1006.3471v1 (2010).

\bibitem{Milton} Milton S. Torikachvili, Sergey L. Bud'ko, Ni Ni, Paul C.
Canfield, Phys. Rev. Lett. \textbf{101}, 057006 (2008).

\bibitem{Alireza} Patricia L. Alireza, Y. T. Chris Ko, Jack Gillett, Chiara
M. Petrone, Jacqueline M. Cole, Gilbert G. Lonzarich, Suchitra E. Sebastian,
J. Phys.: Condens. Matter \textbf{21}, 012208 (2009).

\bibitem{Saha1} S. R. Saha, N. P. Butch, K. Kirshenbaum, and Johnpierre
Paglione, Phys. Rev. Lett. \textbf{103}, 037005 (2009).

\bibitem{Colombier} E. Colombier, S. L. Bud'ko, N. Ni, P. C. Canfield, Phys.
Rev. B \textbf{79}, 224518 (2009).

\bibitem{Kawasaki} S. Kawasaki, T. Tabuchi, X. F. Wang, X. H. Chen and
Guo-qing Zheng, Supercond. Sci. Technol. \textbf{23}, 054004 (2010).

\bibitem{Park} Tuson Park, Eunsung Park, Hanoh Lee, T. Klimczuk, E. D.
Bauer, F. Ronning and J. D. Thompson, J. Phys.: Condens. Matter \textbf{20}
322204 (2008).

\bibitem{Shannon} R. D. Shannon, Acta Cryst., A32 751 (1976).

\bibitem{He} Y He, T Wu, G Wu, Q J Zheng, Y Z Liu, H. Chen, J J Ying, R H
Liu, X F Wang, Y L Xie, Y J Yan, J K Dong, S Y Li and X H Chen, J. Phys.:
Condens. Matter \textbf{22}, 235701 (2010).

\bibitem{Paul} P. C. Canfield and Z. Fisk, Philos. Mag. B \textbf{65}, 1117
(1992).

\bibitem{Saha} S. R. Saha, N. P. Butch, K. Kirshenbaum, and Johnpierre
Paglione, Phys. Rev. B \textbf{79}, 224519 (2009).

\bibitem{Rietica} Hunter B., \textquotedblright Rietica - A visual Rietveld
program\textquotedblright ,International Union of Crystallography Commission
on Powder Diffraction Newsletter No. 20, (Summer) http://www.rietica.org
(1998).

\bibitem{Hidenori} Hidenori Hiramatsu, Takayoshi Katase, Toshio Kamiya,
Masahiro Hirano, and Hideo Hosono, Phys. Rev. B \textbf{80}, 052501 (2009).

\bibitem{Jasche} A. Jesche, N. Caroca-Canales, H. Rosner, H. Borrmann, A.
Ormeci D. Kasinathan, H. H. Klauss, H. Luetkens, R. Khasanov, A. Amato, A.
Hoser, K. Kaneko, C. Krellner, and C. Geibel, Phys. Rev. B \textbf{78},
180504 (2008).

\bibitem{JQ Yan} J.-Q. Yan, A. Kreyssig, S. Nandi, N. Ni, S. L. Bud'ko, A.
Kracher, R. J. McQueeney, R. W. McCallum, T. A. Lograsso, A. I. Goldman, and
P. C. Canfield, Phys. Rev. B \textbf{78}, 024516 (2008).

\bibitem{Prozorov} R. Prozorov, M. A. Tanatar, Bing Shen, Peng Cheng, Hai-Hu
Wen, S. L. Bud'ko, P. C. Canfield, Phys. Rev. B \textbf{82}, 180513(R)
(2010).

\bibitem{Gillett} Jack Gillett, Sitikantha D. Das, Paul Syers, Alison K. T.
Ming, Jose I. Espeso, Chiara M. Petrone, and Suchitra E. Sebastian,
arXiv:1005.1330 (2010).

\bibitem{Deepa} Deepa Kasinathan, Alim Ormeci, Katrin Koch, Ulrich
Burkhardt, Walter Schnelle, Andreas Leithe-Jasper and Helge Rosner, New J.
Phys. \textbf{11} 025023 (2009).

\bibitem{Ren} Zhi Ren, Zengwei Zhu, Shuai Jiang, Xiangfan Xu, Qian Tao, Cao
Wang, Chunmu Feng, Guanghan Cao, and Zhu'an Xu, Phys. Rev. B \textbf{78},
052501 (2008).

\bibitem{Xiao} Y. Xiao, Y. Su, M. Meven, R. Mittal, C. M. N. Kumar, T.
Chatterji, S. Price, J. Persson, N. Kumar, S. K. Dhar, A. Thamizhavel, and
Th. Brueckel, Phys. Rev. B \textbf{80}, 174424 (2009).

\bibitem{Jiang} Shuai Jiang, Yongkang Luo, Zhi Ren, Zengwei Zhu, Cao Wang,
Xiangfan Xu, Qian Tao, Guanghan Cao and Zhu'an Xu, New Journal of Physics 
\textbf{11}, 025007 (2009).

\bibitem{Fisher1} M. E. Fisher, Phil. Mag., \textbf{7}, 1731 (1962).

\bibitem{Jongh} L. J. de Jongh, A. R. Miedema, Adv. Phys., \textbf{23}, 1
(1974).

\bibitem{Taichi1} Taichi Terashima, Nobuyuki Kurita, Akiko Kikkawa, Hiroyuki
S. Suzuki, Takehiko Matsumoto, Keizo Murata, and Shinya Uji, J. Phys. Soc.
Jpn. \textbf{79}, 103706 (2010).

\bibitem{Wiener} T. A. Wiener, I. R. Fisher, S. L. Bud'ko, A. Kracher, and
P. C. Canfield, Phys. Rev. B \textbf{62}, 15056 (2000).

\bibitem{Taichi2} Taichi Terashima, Motoi Kimata, Hidetaka Satsukawa,
Atsushi Harada, Kaori Hazama, Shinya Uji, Hiroyuki S. Suzuki, Takehiko
Matsumoto, and Keizo Murata, J. Phys. Soc. Jpn. \textbf{78} 083701 (2009).

\bibitem{Cho} B. K. Cho, P. C. Canfield, D. C. Johnston, Phys. Rev. Lett., 
\textbf{77}, 163 (1996).

\bibitem{Zhao} J.Zhao, W. Ratcliff, J. W. Lynn, G. F. Chen, J. L. Luo, N. L.
Wang, J. P. Hu, and P. C. Dai, Phys. Rev. B \textbf{78}, 140504 (2008).
\end{thebibliography}
\end{document}